\documentclass
[allcolors=blue]
{optica-article}
\usepackage{xcolor}
\usepackage{siunitx}
\journal{opticajournal} 

\articletype{Research Article}

\DeclareMathOperator{\sinc}{sinc}

\begin{document}
\title{High-purity frequency-degenerate photon pair generation via cascaded SFG/SPDC in thin film lithium niobate}

\author{Olivia Hefti,\authormark{1,2,*} 
Marco Clementi,\authormark{3}
Enrico Melani,\authormark{3,$\dagger$} 
Jean-Etienne Tremblay,\authormark{1,$\ddagger$} 
Andrea Volpini,\authormark{1}
Yesim Koyaz,\authormark{2}
Homa Zarebidaki,\authormark{1}
Ivan Prieto,\authormark{1,$\S$}
Olivier Dubochet,\authormark{1}
Daniele Bajoni,\authormark{4}
Charles Caër,\authormark{1}
Hamed Sattari,\authormark{1,¶}
Camille-Sophie Brès,\authormark{2}
Matteo Galli,\authormark{3}
 and Davide Grassani,\authormark{1}}

\address{\authormark{1}Centre suisse d'électronique et de microtechnique, 2000 Neuchâtel, Switzerland\\
\authormark{2}Photonic Systems Laboratory, École Polytechnique Fédérale de Lausanne, 1015 Lausanne, Switzerland\\
\authormark{3}Dipartimento di Fisica “A. Volta”, Università di Pavia, 27100 Pavia, Italy \\
\authormark{4}Dipartimento di Ingegneria Industriale e dell’Informazione, Università di Pavia, 27100 Pavia, Italy
}

\email{\authormark{*}olivia.hefti@csem.ch} 
\email{\authormark{$\dagger$}Now at: Department of Electronic Systems, Norwegian University of Science and Technology, 7491 Trondheim, Norway \\}
\email{\authormark{$\ddagger$}Now at: {Enlightra, 1020 Renens, Switzerland}\\}
\email{\authormark{$\S$}Now at: International Iberian Nanotechnology Laboratory (INL), 4715-330 Braga, Portugal \\}

\email{\authormark{¶}Now at: CCRAFT Thin-Film Lithium Niobate Photonic Chip Foundry, 2000 Neuchâtel, Switzerland\\}

\begin{abstract*}  
Frequency-degenerate photon pairs generated using nonlinear photonic integrated devices are a crucial resource for scalable quantum information processing and metrology. 
However, their realization is hindered by unwanted parametric processes occurring within the same phase matching band, which degrade the signal-to-noise ratio and reduce the purity of the associated quantum states. 
Here, we propose a dual-pump scheme to produce frequency-degenerate photon pairs, based on cascaded sum-frequency generation and spontaneous parametric down-conversion occurring within a single waveguide, while strongly suppressing parasitic photon pair generation from single-pump processes.
This approach significantly simplifies the design compared to microresonator-based methods and enables both pumping and collection of photon pairs entirely in the telecom band. 
We experimentally validate the concept in a layer-poled thin film lithium niobate waveguide, achieving frequency-degenerate photon pair generation with a brightness of \SI{1.0(3)e5}{\hertz \per \nm \per \square \milli \watt } and a 40 dB suppression of unwanted single-pump processes.

\end{abstract*}

\section{Introduction}

Frequency-degenerate photon pairs, which underlie single-mode squeezed states, are essential resources in continuous variable (CV) quantum information. 
They enable a wide range of applications, including quantum computing \cite{vernon2019scalable}, quantum error correction \cite{braunstein1998error}, quantum key distribution \cite{gottesman2001secure} and metrology measurements beyond the standard quantum limit \cite{giovannetti2004quantum}. 
In integrated photonics, photon pairs can be generated through two nonlinear optical processes: spontaneous parametric down conversion (SPDC), relying on second-order nonlinear susceptibility ($\chi^{(2)}$) \cite{lu2021advances,caspani2017integrated} in materials such as lithium niobate (LN) \cite{zhao2020high,xue2021ultrabright}, aluminium gallium arsenide (AlGaAs) \cite{placke2023telecom}, aluminium nitride (AlN) \cite{guo2017parametric} and silicon carbide \cite{shi2025spontaneous}; or spontaneous four-wave mixing (SFWM), exploiting third-order nonlinearities ($\chi^{(3)}$) in platforms like silicon on insulator (SOI) \cite{guo2014impact,grassani2015micrometer,silverstone2015qubit}, silicon nitride (Si$_3$N$_4$)\cite{lu2019chip,ramelow2015silicon} and high-index doped silica (Hydex) \cite{reimer2016generation}. 
Since $\chi^{(3)}$ nonlinear processes are intrinsically less efficient than their $\chi^{(2)}$ counterparts, they typically require field enhancement in optical cavities to boost the nonlinear conversion efficiency. 
In particular, frequency-degenerate photon pairs based on $\chi^{(3)}$ processes are commonly generated via dual-pump spontaneous four-wave mixing (DP-SFWM) in microresonators \cite{guo2014telecom, vernon2019scalable}. 
However, dual-pumping a single microresonator inherently suffers from parasitic noise due to single-pump (SP-) SFWM, which decrease the purity of the generated quantum states. 
It has been shown that these unwanted SP-SFWM contributions can be suppressed by detuning the pumps from resonance  \cite{zhao2020near} or by leveraging two resonators that are either strongly linearly coupled \cite{zhang2021squeezed}, or linearly uncoupled but nonlinearly coupled \cite{sabattoli2021suppression}. 
Such methods, although effective in suppressing SP processes, result in compromising bandwidth and efficiency. 
More recently, alternative methods based on a resonant interferometric coupler \cite{viola2024squeezing} and on a single nano-corrugated photonic crystal ring (PhCR) \cite{ulanov2025quadrature} have been proposed to suppress SP processes. 

In $\chi^{(2)}$ waveguides, photon pairs in the telecom band are typically produced either via SPDC using a pump in the near-visible \cite{zhao2020high}, or through second harmonic generation (SHG) followed by SPDC \cite{park2024single,elkus2020quantum}. 
The latter approach is particularly convenient when pumping in the telecom-band  is required, for instance when the device is integrated into a fiber-optic network \cite{furukawa2007tunable, hu2011polarization} or in applications that require coherent mixing of the generated quantum state with a local oscillator derived from the same pump \cite{arge2024demonstration}. 
However, such scheme demands strong suppression of the residual pump light after the SHG stage \cite{park2024single}, to isolate the frequency-degenerate photon pairs.

Here we propose and experimentally demonstrate a method to suppress the generation of parasitic photons from SP parametric processes by leveraging cascaded sum-frequency generation (SFG) and SPDC (cSFG/SPDC) in the same thin film lithium niobate (TFLN) waveguide. Owing to the high $\chi^{(2)}$ and low propagation losses of LN, the cSFG/SPDC process achieves significantly higher efficiency than SFWM in $\chi^{(3)}$ platforms, and does not require additional filtering stage. We report an on-chip brightness as high as \num{1.0(3)e5} \unit{\hertz \per \nm \per \square \milli \watt} and show that the SP parasitic processes produce a negligible amount of photon pairs at the degenerate point, i.e. $10^4$ times lower (40 dB suppression) than the desired frequency-degenerate photon pairs from the DP process. We also investigate the impact of LN spontaneous Raman scattering on photon production.

\begin{figure}[htbp]
\centering\includegraphics[width=12 cm]{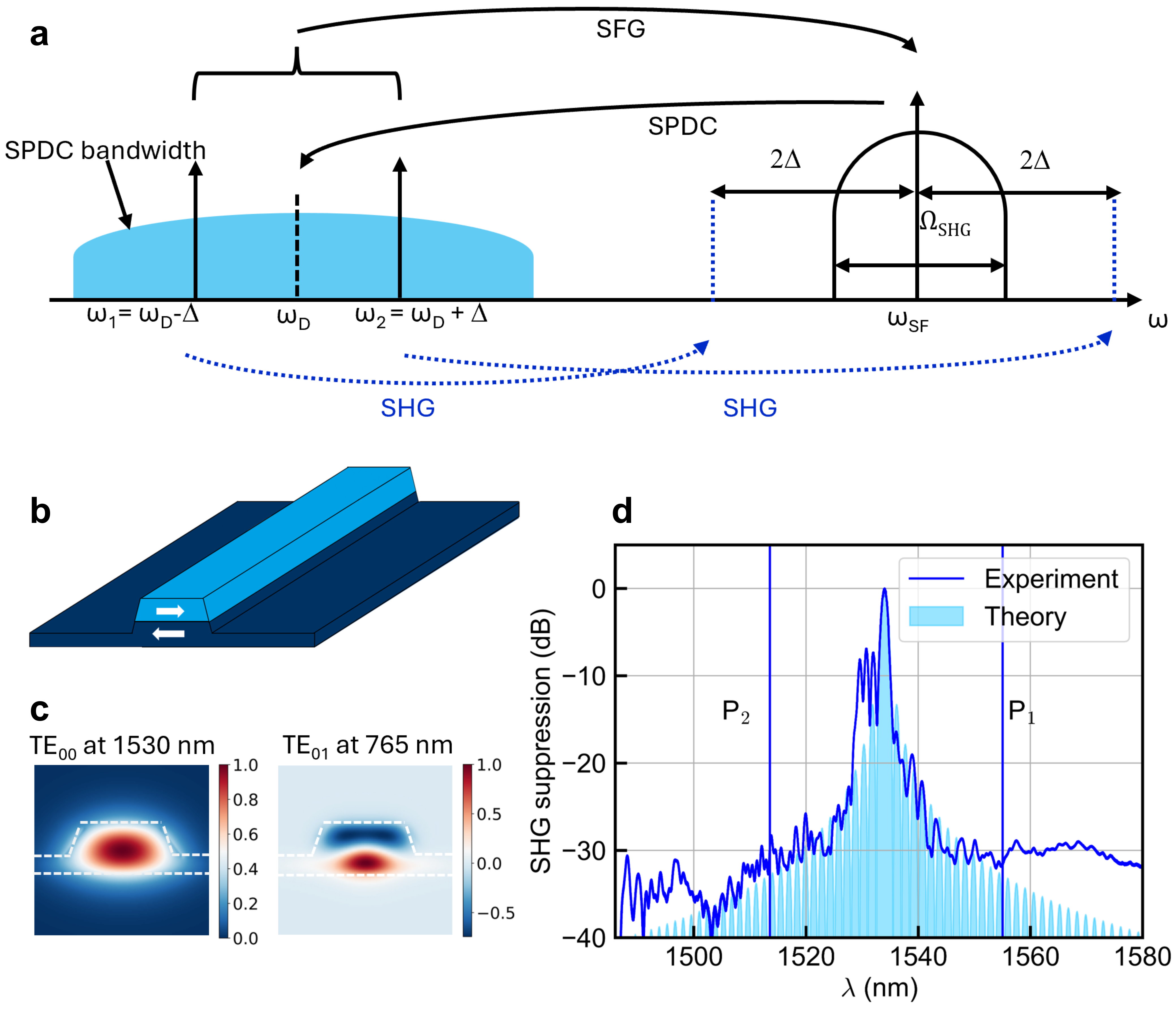}
\caption{(a) Cascaded SFG/SPDC process: dual-pump scheme where two input pumps at $\omega_1$ and $\omega_2$ create SFG at $\omega_\mathrm{SF} =\omega_1 + \omega_2$ which acts as the pump for frequency-degenerate SPDC. 
$\omega_\mathrm{SF}$ is centered on the SHG phase matching frequency. 
The single-pump SHG processes are suppressed as they fall outside of the phase matching bandwidth of SHG, $\Omega_{\mathrm{SHG}}$. 
(b) Ideal layer-poling for the MPM between the (c) fundamental TE$_{00}$ mode in the telecom band and the TE$_{01}$ mode in the near-visible. 
The LN ferroelectric domains at the bottom of the waveguide (dark blue) are reversed compared to the ones on top (light blue). The boundary sit at half the waveguide height. (c) Shows the real part of the transverse (TE) component of the electrical fields, normalized such that the maximum over the cross-section is equal to 1. (d) Theoretical and experimental SHG spectrum, from which we infer the SHG suppression away from phase matching.}
\label{concept_setup}
\end{figure}

\section{Theoretical background}
We start by introducing the nonlinear processes of SFG, SHG and SPDC in a TFLN waveguide. 
Detailed derivations of the coupled wave equations (CWEs) for both SFG and the cascaded SFG/SPDC process in the frequency-degenerate case are provided in Sections 1 and 2 of the supplementary material.
SFG is a process mediated by $\chi^{(2)}$, whose value is intrinsic to non-centrosymmetric materials. 
In SFG, two input waves at frequencies $\omega_1$ and $\omega_2$ (referred to as the pump waves) interact to generate a new wave at the sum frequency $\omega_{\mathrm{SF}}=\omega_1 + \omega_2$, in accordance with energy conservation. 
For efficient SFG, momentum conservation, also known as the phase-matching condition, must be satisfied as well. 
This is expressed by the condition: $\Delta k_{\mathrm{SF}} = k_{\mathrm{1}}(\omega_1) + k_{\mathrm{2}}(\omega_2)-k_{\mathrm{SF}}(\omega_{\mathrm{SF}})=0$, where $k_j(\omega_j)$ with $j = 1,2,\mathrm{SF}$ denotes the wavenumber of each interacting optical mode.
Due to material and waveguide geometry dispersion, phase matching between fundamental transverse modes at spectrally distinct frequencies is generally not satisfied automatically.
A common approach to overcome this limitation is quasi-phase matching (QPM), which enables SFG between identical optical modes by periodically inverting the material’s ferroelectric domains with a period $\Lambda$. 
This introduces a grating wavevector into the phase-matching condition, modifying it (at first order) to: $\Delta k_{\mathrm{SF}} = k_{\mathrm{1}}(\omega_1) + k_{\mathrm{2}}(\omega_2)-k_{\mathrm{SF, QPM}}(\omega_{\mathrm{SF}}) - \frac{2\pi}{\Lambda}=0$ \cite{boyd2008nonlinear}. 
Alternatively, modal phase matching can be used, where the pump waves propagate in fundamental modes while the sum-frequency (SF) wave occupies a higher-order mode, typically associated with a lower effective index. 
In this case, phase matching is achieved intrinsically by satisfying $\Delta k_{\mathrm{SF}} = 0$ \cite{hefti2025fabrication}. 
Importantly, the phase matching condition and the length of the waveguide, $L$, determine the spectral bandwidth of the SFG process, as can be seen from the expression of the SF power:

\begin{equation}
    P_{\mathrm{SF}}(L) = \kappa^2_{\mathrm{SFG,SF}} P_1 P_2 L^2 \sinc^2\Big(\frac{\Delta k_{\mathrm{SF}} L}{2}\Big).
\end{equation}
Here $P_1$ and $P_2$ are the powers of the pumps and $\kappa_{\mathrm{SFG,SF}}$ is the nonlinear coupling coefficient, defined in the supplementary material. 
This expression assumes the undepleted pump approximation, i.e., the pump powers remain constant along the propagation direction: $\frac{\partial P_1}{\partial z} = \frac{\partial P_2}{\partial z} =0$. 
We define the peak normalized SFG conversion efficiency as:
\begin{equation}
  \eta_{\mathrm{SFG}} =   
  \frac{P_{\mathrm{SF}} }{P_\mathrm{1} P_\mathrm{2} L^2 } =   
  \kappa^2_{\mathrm{SFG,SF}}.
\end{equation}

\noindent
SHG is a particular case of SFG in which the two pump photons are degenerate, i.e. a single optical pump is used. In this process, two photons at frequency $\omega$ combine to generate a new photon at the SH frequency $\omega_{\mathrm{SH}} = 2\omega$. The phase-matching condition for efficient SHG is given by: $\Delta k_{\mathrm{SH}} = 2k_\mathrm{p}  -k_{\mathrm{SH}}$. Under the undepleted pump approximation, the power of the SH is given by

\begin{equation}
    P_{\mathrm{SH}}(L) = \kappa^2_{\mathrm{SHG}} P_\mathrm{p}^2 L^2 \sinc^2\Big(\frac{\Delta k_{\mathrm{SH}} L}{2}\Big),
    \label{eq:SHG_sinc}
\end{equation}
and the conversion efficiency is defined as $\eta_\mathrm{SHG}=\frac{P_\mathrm{SH}}{P_\mathrm{p}^2L^2}$. The full-width at half maximum (FWHM) SHG bandwidth, $\Omega_\mathrm{SHG}$, is inversely proportional to $L$. In TFLN waveguides, with $L\sim 5$ mm and a pump in the telecom band, the SHG bandwdith $\frac{\Omega_\mathrm{SHG}}{2\pi}$ typically lies in the 250-750 GHz range.

In SPDC, a pump photon at $\omega_{\mathrm{p,SPDC}}$ spontaneously annihilates by emitting a pair of lower-energy photons at $\omega_1$ and $\omega_2$ such that $\omega_{\mathrm{p,SPDC}} = \omega_1 + \omega_2$. SPDC is described by the semi-classical Hamiltonian \cite{u2005generation}:
\begin{equation}
    \hat{H} = \iint f(\omega_1,\omega_2)\hat{a_1}^\dagger\hat{a_2}^\dagger \mathrm{d}\omega_1 \mathrm{d}\omega_2 + \mathrm{h.c.},
\end{equation}
where $\hat{a_i}^\dagger$ are the creation operators and $f(\omega_1,\omega_2)= \phi(\omega_1,\omega_2)\alpha(\omega_1,\omega_2)$ is the joint spectral amplitude of the biphoton state, which is a product of the pump envelope function, $\alpha(\omega_1,\omega_2)$, and of the phase matching function, $\phi(\omega_1,\omega_2)$. The term $\alpha(\omega_1,\omega_2)$ depends on the pump spectrum and it ensures energy conservation, while $\phi(\omega_1,\omega_2)$ reflects the impact of the waveguide dispersion on the nonlinear process for each pair at $\omega_1$ and $\omega_2$.

\section{Results}

\subsection{Implementation}
The scheme we use is illustrated in Fig.~\ref{concept_setup} a). 
Two pump lasers at frequencies $\omega_1=\omega_\mathrm{D}-\Delta$ and $\omega_2=\omega_\mathrm{D}+\Delta$ in the telecom band are symmetrically placed around the SHG phase matched frequency, $\omega_D$, to ensure phase matching of the SFG process. 
The pumps generate the sum frequency at $\omega_\mathrm{SF} = \omega_1+\omega_2$, which acts as the pump for SPDC, producing frequency-degenerate photon pairs back in the telecom band around $\omega_\mathrm{D} = (\omega_1+\omega_2)/2$. 
Importantly, by placing the two pumps outside $\Omega_\mathrm{SHG}$, the photon pairs emitted from cSHG/SPDC can be suppressed. 

To enable the nonlinear processes, we exploit layer-poled modal phase matching (MPM) in a TFLN waveguide, which requires an inversion of the ferroelectric domain between the bottom and the top of the waveguide, as shown in Fig.~\ref{concept_setup} b) \cite{hefti2025fabrication, shi2024efficient}. 
This MPM mechanism enables efficient frequency conversion between the fundamental TE$_{00}$ mode in the telecom band and the TE$_{01}$ mode in the near-visible, shown in Fig.~\ref{concept_setup} c), showing good efficiency and less sensitivity to fabrication imperfections \cite{hefti2025fabrication, shi2024efficient}. The process would be highly inefficient in the absence of layer poling, owing to the symmetry of the modes involved. Note that our scheme can also be used with cSHG/SPDC based on QPM.

The experimental chip is fabricated using CSEM's open-access X-cut TFLN foundry process. The cross-section has a target total thickness of 600 nm and a slab thickness of  \SI{200}{\nano\meter}. We use an air-cladded waveguide which is \SI{1.0}{\micro\meter} wide, 4.45 mm long, and features inverse tapers with two etching steps at both input and output \cite{he2019low}. The SHG phase matching wavelength occurs at  \SI{1534}{\nano\meter} and the SHG spectrum is shown in Fig.~\ref{concept_setup} e). The waveguide has been poled after etching to enable the previously mentioned MPM process and has been characterized for the classical process of cascaded SHG and stimulated DFG (cSHG/DFG) in previous works \cite{hefti2025fabrication}. To estimate the SHG suppression as a function of the detuning from the phase matching wavelength, we simulate the dispersion of the waveguide and use Eq. \ref{eq:SHG_sinc} to obtain the theoretical SHG spectrum, which is shown in Fig. ~\ref{concept_setup} d) together with the experimental spectrum. The distortion of the latter is likely due to film thickness inhomogeneity along the waveguide. We see that placing the pumps at 1513.56 and  \SI{1555.05}{\nano\meter}, indicated with vertical lines, the theory predicts >30 dB of suppression, which is consistent with our measurement in Fig. ~\ref{concept_setup} d). It is important to notice  that, in Fig. ~\ref{concept_setup} d), the SHG signal is assumed to be generated by a pump power $P$ for all the wavelengths. However, in the SFG process, we consider $P_1 = P_2 = P/2$. Therefore, while the SFG and SHG signals have the same amplitude at the phase-matching wavelentgh (1534 nm), the SHG from $P_1$ and $P_2$ should be further reduced by a factor 4 (-6 dB) with respect to the plot, leading to a total theoretical suppression of >36 dB.

\subsection{Spontaneous Raman scattering and SPDC spectrum}
\begin{figure}[htbp]
\centering\includegraphics[width=14 cm]{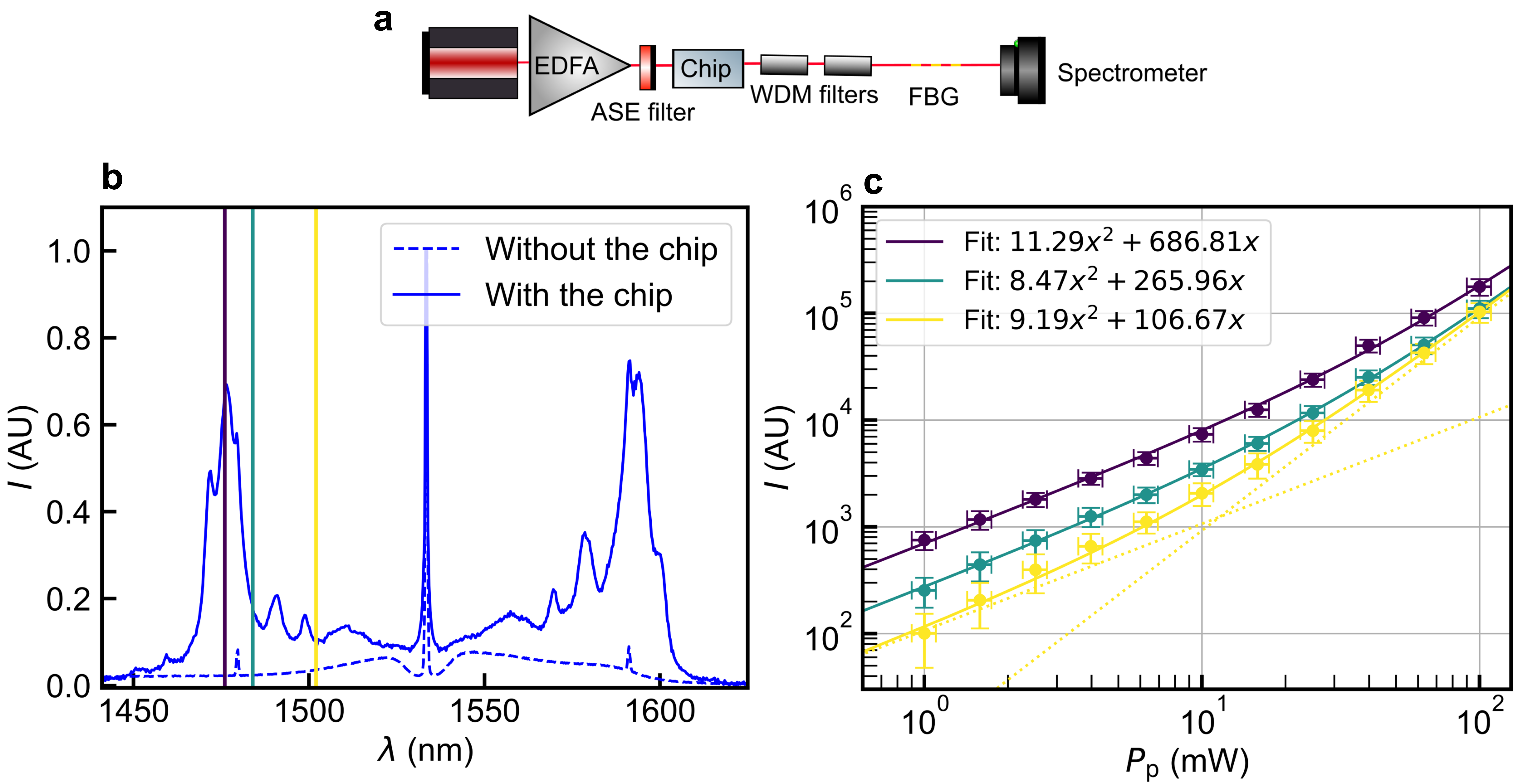}
\caption{(a) Setup used to measure the emission spectrum from the chip. EDFA: erbium doped fiber amplifier, ASE filter: amplified spontaneous emission filter, WDM filters: wavelength division multiplexer, FBG: fiber bragg grating. (b) Emission spectrum from the TFLN chip centered around the pump wavelength (solid) and residual contribution from the fibers, obtained bypassing the chip (dashed). The highest peaks indicate the Raman shifts associated with the mode 251 1A1. The spectrometer intensity at the wavelengths indicated with vertical lines are recorded while increasing the pump power and shown in (c). The experimental data are fitted on a linear scale with the model $ax^2 + bx$, which includes a quadratic term (from cSHG/SPDC) and a linear term (from Raman scattering), and shown in panel c). The two dashed lines are a guide to the eye representing pure linear and pure quadratic scaling, illustrating the transition from a Raman-dominated to an SPDC-dominated regime. The Raman coefficient is 7 times higher at the 251 1A1 mode position than at 1502 nm.} \label{Raman}
\end{figure}

Although our scheme mainly addresses the suppression of photons coming from the single pump cSHG/SPDC process, in this section, we examine whether other processes competing with SPDC produce photons in the telecom band, and must therefore be taken into account when selecting optimal parameters for the dual-pump experiment. 
We start using the setup shown in Fig.~\ref{Raman} a). A single pump laser (Santec TSL-570), amplified with an erbium-doped fiber amplifier (EDFA), is filtered with a bandpass filter (Semrock NIR01-1535/3-25) to suppress its amplified spontaneous emission (ASE) and side-coupled to the waveguide using an aspheric lens that matches the numerical aperture of the waveguide taper. We obtain 5 dB of insertion loss per facet. At the output, the residual pump is suppressed from the collected light using two wavelength division multiplexers (Opneti DWDM-1-1-C55-900-1-1-FA) with an extinction ratio (ER) of 12 dB each and a fiber Bragg grating (OEDWDM-0125) with an extinction ratio of 31 dB. 
Then, light is sent to a high-sensitivity infrared spectrometer (Princeton instruments, Spectra Pro 2500i, InGaAs detector). 

The spectrum in Fig.~\ref{Raman} b) shows a peak at  \SI{1534}{\nano\meter}, corresponding to the residual pump, and a series of peaks symmetrically centered around it. The shifts between these peaks and the pump match very well the Raman shift in LN measured in a $X$ ($ZZ$, $ZY$) configuration \cite{gorelik2019raman}, i.e. when light propagates along an axis perpendicular to the $Z$ axis of the crystal, like in the case of the $X$-cut LN waveguide used here. 
From \cite{gorelik2019raman}, we can identify the peaks with a shift of $\pm$ 251 \unit{\per \cm}, $\pm$ 238 \unit{\per \cm}, $\pm$ 151 \unit{\per \cm}, with the vibrational modes: 1A1 TO, 2E TO, 1E TO, respectively. 
We confirm that spontaneous Raman scattering occurs in the chip by taking the spectrum of the light passing through the same setup but without the chip (dashed trace). Here we see a broad Raman scattering spectrum from the fibers and three small peaks corresponding to different orders of diffraction of the pump by the spectrometer’s grating, while the strong Raman signatures visible in the other spectrum have disappeared.

The detected signal on the spectrum consists in a contribution from Raman scattering and a contribution from cSHG/SPDC. Raman scattering scales linearly with the pump power and it is wavelength dependent, while cSHG/SPDC scales quadratically with the pump power and is broadband (>100 nm); thus we expect its contribution to be constant at different wavelengths. 

The spectrometer intensity at the wavelengths indicated with vertical lines in Fig.~\ref{Raman} b) (1475, 1483 and 1502 nm) are recorded while increasing the pump power and shown in Fig.~\ref{Raman} c). The experimental data are fitted on a linear scale using the polynomial model $ax^2 + bx$, where the quadratic and linear terms correspond to the contributions from cSHG/SPDC and from Raman scattering, respectively. The fitted coefficients are presented in Fig.~\ref{Raman} c). Notably, the Raman coefficient is approximately 7 times larger at the 251 1A1 mode position than at 1502 nm. 
To minimize the contribution from spontaneous Raman scattering in the following experiments, we collected photon pairs with a detuning in absolute value $<50$ nm from the pump wavelengths and $>5$ nm to have the pumps outside $\Omega_{\mathrm{SHG}}$.

\subsection{Performance of photon pair source in dual-pump configuration}
We now assess the performance of our frequency-degenerate photon source in DP configuration. 
The setup we use is shown in Fig.~\ref{Experiment} a).
We combine two pumps using a coarse wavelength division multiplexer (CWDM) (FS 18ch dual fiber CWDM mux/demux), which also suppresses the ASE of the sources. At the output, the collected light is filtered by a tunable waveshaper (Finisar WS-16000A) configured as a squared notch filter with an adjustable bandwidth and 3 CWDM stages to effectively reject any residual pump. The 1530 nm output channel of the CWDM is sent to a 50/50 fiber beam-splitter (BS) connected to two polarization controllers (PC) followed by two superconducting nanowire single photon detectors (SNSPD). 
The photon arrival times are recorded and correlated by a time tagger (qutools QuTAG time-to-digital converter). The time jitter (FWHM) of the SNSPDs is $\sim 50$ ps. The pump wavelengths are chosen at 1513.56 and 1555.05 nm, to lie outside $\Omega_{\mathrm{SHG}}$, as can be verified on the SHG spectrum shown in Fig.~\ref{concept_setup} e), and to minimize Raman scattering at 1534 nm, where the frequency-degenerate photon pairs are generated. 
We start by setting the detection bandwidth of the tunable filter to 1 nm. Fig.~\ref{Experiment} b) shows the pair generation rate (PGR), defined in the supplementary material, by increasing the power of the pump at 1555.05 nm and keeping the other constant at 0.9 mW (on-chip). The left (right) axis shows the estimated on-chip (detected) PGR. As expected, we observe a linear dependence, as SFG is proportional to the product of the pump powers and that SPDC grows linearly with the SFG power. 
From the estimated PGR, we derive an on-chip brightness of $B=\frac{PGR}{4 \Delta \lambda P_\mathrm{1} P_\mathrm{2}} =$ \num{1.0(3)e5} \unit{\hertz \per \nm \per \square \milli \watt }, where $\Delta \lambda$ is the SPDC measurement bandwidth. 

Fig.~\ref{Experiment} c) shows the coincidence-to-accidental ratio (CAR), defined in the supplementary material, associated with the measurements in Fig.~\ref{Experiment} b). 
The maximal CAR of 372$\pm$ 5, is obtained when the two pumps have similar power (here $P_{2}=$ 1.7 mW) as it minimizes the single-pump SHG contributions relative to SFG. Above this value, the generation of multi-photon pairs starts to dominate. At lower power, the CAR is limited by noise photons coming from detector dark counts, Raman scattering, and parasitic SP processes. As we will show in the next section, Raman scattering turns out to be the dominating factor. 

In Fig.~\ref{Experiment} d), we show that by narrowing the detection bandwidth $\Delta \lambda$, thus suppressing the accidental counts from Raman scattering more effectively, the CAR can be significantly increased up to $>$2000 for $\Delta \lambda$= 0.1 nm. This is a trade-off as reducing $\Delta \lambda$ also implies a reduction of the PGR.

\subsection{Suppression of parasitic photons from single-pump parametric processes}
We now demonstrate the suppression of SP parametric processes by comparing the number of coincidental counts from cSHG/SPDC where only one photon from each pair is emitted in the detection bandwidth of the frequency-degenerate photon pairs ($\omega_\mathrm{D}\pm \frac{c \pi\Delta \lambda}{\lambda^2}$ where $\lambda= 1534$ nm), with the number of coincidental counts from cSFG/SPDC processes where both photons from the pairs fall within this detection bandwidth. To reduce the integration time of the measurement, we set $\Delta \lambda = 2 $ nm.

\begin{figure}[htbp]
\centering\includegraphics[width=15 cm]{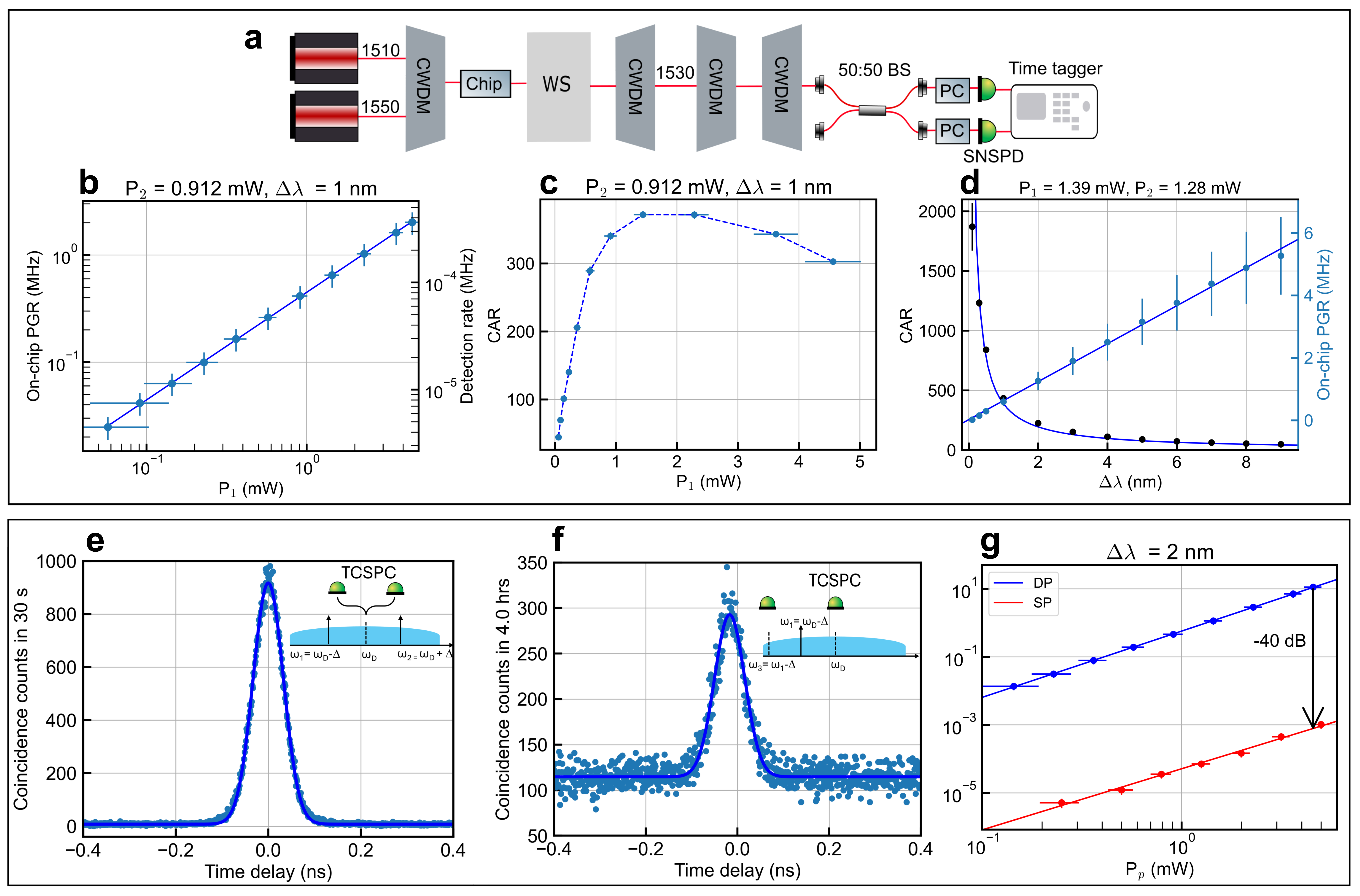}
\caption{(a) Setup to measure the frequency-degenerate photon pairs from cSFG/SPDC in DP experiment. WS: tunable waveshaper in a notch filter configuration, CWDM: coarse-wavelength division multiplexer, BS: beam splitter, PC: polarization controller, SNSPD: superconducting nanowire single photon detector. (b) Estimated on-chip (detected) pair generation rate data on the left (right) axis, fitted with a linear regression in a log-log scale (solid line), scaling the power of one pump power and keeping the other constant at 0.17 mW. (c) CAR for the same power scaling. (d) CAR (on-chip PGR) as a function of the detection bandwidth on the left (right) axis. The CAR is fitted using $ax^{-1}$ with $a=388$, while the PGR is fit with a linear model $ax$. (e) Time-correlation histogram for frequency-degenerate photon pairs at $\omega_\mathrm{D}$ from cSFG/SPDC, measured as shown in the inset. TCSPC: time-correlated single photon counting
(f) Time-correlation histogram for non-degenerate photon pairs from cSHG/SPDC, with one photon from each pair at $\omega_\mathrm{D}$, measured as shown in the inset. (g) PGR of the DP processes (linear fit: $1.95x -2$) and PGR of the SP processes (linear fit: $1.8x -43$), showing a relative suppression of 40 dB consistently for various pump power. }
\label{Experiment}
\end{figure}

First, we use the configuration shown on the inset in Fig.~\ref{Experiment} e), with two pumps at $\omega_1$ and $\omega_2$ and the frequency-degenerate photon pairs from cSFG/SPDC collected at $\omega_\mathrm{D}$. Fig.~\ref{Experiment} e) shows the associated time-correlation histogram, measured with an integration time of 30 s, from which we infer an on-chip PGR of 11.26 MHz. 

Then, we use a single pump at 1555.05 nm with the same power as in Fig.~\ref{Experiment} e) and an integration time of 4 hours (14400 s). As the inset in Fig.~\ref{Experiment} f) shows, we assess the contribution from the individual cSHG/SPDC processes by looking at the coincidence rate of photon pairs at $\omega_\mathrm{D}$ and $\omega_3 = \omega_1-\Delta$, where $\Delta = |\omega_1 - \omega_\mathrm{D}|$. The associated time-correlation histogram is shown in Fig.~\ref{Experiment} f) and corresponds to an on-chip PGR of 1.1 kHz. By definition, PGR is rescaled by the integration time, so we can directly compare the PGR from the SP and DP experiments and show that the SP parasitic processes are suppressed by 40 dB. Note that the photon pairs measured in Fig.~\ref{Experiment} f) could also be produced by DP-SFWM. To exclude this possibility, we compare the efficiency of the classical counterparts of these processes, i.e. cSHG/DFG and stimulated FWM. We estimate cSHG/DFG (detuned by 20 nm from phase matching as in the DP experiment) to be 6 orders of magnitude more efficient than stimulated FWM in our waveguides, see section 2 in the supplementary materials for more details.

\section{Discussion and conclusions}

The photon PGR in this work is limited by the nonlinear conversion efficiency of the waveguide, which we estimate to be $\eta_\mathrm{SHG}=250$ \unit{\percent \per\watt \per\square\cm}. However, this value does not represent an intrinsic limit of layer-poled MPM, as SHG efficiency as high as 4615 \unit{\percent \per\watt \per\square\cm} has been experimentally demonstrated \cite{shi2024efficient}. The main limitations to the conversion efficiency in this work are the shallow poling depth (less than half the waveguide height) and the inhomogeneous poling along the waveguide length caused by the deep comb-like design of our electrodes. This results in an effective nonlinear interaction length shorter than the physical waveguide length; a more detailed analysis is provided in \cite{hefti2025fabrication}. In our process, the poling depth is limited by the slab thickness, therefore an etching depth below 400 nm would allow for a thicker poling section. Furthermore, future electrode designs could improve poling uniformity by employing shorter periods, larger duty cycle, or shallower or continuous electrodes, as demonstrated in \cite{shi2024efficient}.

The limited second-order conversion efficiency implies that the rate of noisy photons generated by Raman scattering are comparable, at low pump power, to the photon pairs produced by the cascaded process, as observed in photon coincidence measurements from TFLN periodically poled waveguides \cite{elkus2020quantum}. 
Indeed, cSFG/DFG is proportional to the product of the pumps and scales approximately as $L^4$,  $\eta^2_\mathrm{SHG}$ under the low pump depletion approximation \cite{xu2004cascaded} (see Supplementary  section 2). Consequently, by increasing the SHG conversion efficiency to $\sim$ \SI{4500}{\percent \per\watt \per\square\cm}, the ratio of PGR to Raman scattering noise could be improved by a factor of $\sim$ 300. Also, at high pump power the detrimental effect of Raman becomes negligible, as shown in Fig. \ref{Raman} c).

Even if our sample shows sub-optimal conversion efficiency, thanks to the high $\chi^{(2)}$ of LN and to the low propagation loss, the cascaded process offers performance comparable, or even higher, than the state-of-the art source of frequency-degenerate photon pairs  based on DP-SFWM and exploiting $\chi^{(3)}$ in resonant structures. For example, \cite{sabattoli2021suppression} uses two linearly uncoupled SOI microresonators and reports an on-chip photon pair generation rate of 62 kHz using 0.95 mW. We estimate their source brightness to be on the order of \num{1e7} \unit{\hertz \per \nm \per \square \milli \watt }, and they achieve a 37 dB suppression of SP-SFWM. In comparison, our system achieves a pair generation rate of 4 MHz, under the same pump powers. We report a suppression of SP parasitic processes by 40 dB and a source spectral brightness two orders of magnitude lower, attributed to our significantly broader collection bandwidth. 
In \cite{zhang2021squeezed} the authors employ two strongly coupled Si$_3$N$_4$ microresonators and leverage an avoided mode crossing to suppress the resonance of SP parasitic processes. Based on the reported relative field suppression factor $\mathcal{R}=\frac{F_\mathrm{DP,SWFM}}{F_\mathrm{SP,SWFM}}\approx 60$ (see Eq. S10 in the supplementary of \cite{zhang2021squeezed}), where the $F_i$ are the overall resonant field enhancements associated with the SP- and DP-SFWM processes, we estimate a suppression of the parasitic process of $\propto \mathcal{R}^2 =$ 35.5 dB, given that the FWM conversion efficiency in resonators scales with the square of the overall field enhancement $F_i$ \cite{wu2022optimization}.

One promising application for the cSFG/SPDC process is the generation of degenerate squeezed vacuum states. Some applications, such as time-domain multiplexed continuous-variable (CV) quantum computing, require broadband squeezed light and photons generated in a highly-pure quantum state. Indeed, a large squeezing bandwidth enables short time bins, and ultimately a fast computer \cite{jankowski2021dispersion}. 
In microresonator-based approaches, high quantum state purity is achieved by restricting the squeezing to discrete cavity resonances, which act as spectral filters. However, this comes at the cost of reduced bandwidth: high-Q (quality factor) resonators offer narrow linewidths that improve mode purity, but inherently limit the bandwidth over which squeezing occurs. Furthermore, although increasing the conversion efficiency, resonant structures induce additional intrinsic losses due a finite coupling between the ring and the bus-waveguide, ultimately decreasing the squeezing level. Overall, the resonance properties of microresonators entail a fundamental trade-off between the squeezing bandwidth and the rate at which high-purity squeezed states can be generated. In straight waveguides, high state purity can be achieved by engineering the waveguide dispersion to produce a separable joint spectral amplitude, resulting in indistinguishable photon pairs \cite{xin2022spectrally}. When combined with a pulsed laser whose bandwidth exceeds that of the SPDC process (typically using femtosecond-duration pulses), this approach enables the generation of squeezed states with broad bandwidth \cite{jankowski2021dispersion}. 

In conclusion, we have demonstrated a novel method to effectively suppress noisy photons arising from SP parametric processes, achieving a suppression of 40 dB, which significantly improves the quality of frequency-degenerate photon pairs in photonic integrated devices. Our approach is based on cascaded second order nonlinearities in a layer-poled TFLN waveguide, though alternative phase matching schemes may also be employed. The proposed scheme preserves the same input and output frequencies as DP-SFWM, but offers 100 times higher generation and a significantly simpler single-pass configuration, while keeping the compatibility with well-established telecom band infrastructure.

\begin{backmatter}
\bmsection{Funding}
This project is funded by the Swiss National Science Foundation (SNSF) (194693), Italian Ministry of Education (MUR) PNRR project PE0000023-NQSTI.


\bmsection{Disclosures}
The authors declare no conflicts of interest.

\bmsection{Data Availability Statement}
The data that support the findings of this study are available from the corresponding author upon reasonable request.

\end{backmatter}




\bibliography{references}

\end{document}